\documentstyle[twoside,fleqn,espcrc2,epsf]{article}

\newcommand{\nc}{\newcommand}
\nc{\aleq}{\mbox{}_{\textstyle \sim}^{\textstyle < }}
\nc{\ageq}{\mbox{}_{\textstyle \sim}^{\textstyle > }}
\def\gtap{\;\raisebox{-.4ex}{\rlap{$\sim$}} \raisebox{.4ex}{$>$}\;}  

\newcommand{\AmS}{{\protect\the\textfont2
  A\kern-.1667em\lower.5ex\hbox{M}\kern-.125emS}}
\title{Topology of full QCD
\thanks{Talk by Ph. de Forcrand, {\bf forcrand@scsc.ethz.ch}}}

\author{
        Philippe de Forcrand\address{Swiss Center for Scientific Computing, 
        ETH-Zentrum, CH-8092 Z\"urich, Switzerland},
Margarita Garc\'{\i}a P\'erez\address{Institut f. Theoretische Physik, 
University of Heidelberg, D-69120 Heidelberg, Germany},
James E. Hetrick\address{Physics Department, 
University of the Pacific, Stockton, CA 95211} and
Ion-Olimpiu  Stamatescu$^{\rm b,}$\address{FEST, Schmeilweg 5, 
D-69118 Heidelberg, Germany}
} 
\begin{document}
\begin{abstract}
We study topological properties of SU(3) gauge theory using improved cooling.
In the absence of fermions, we measure a topological susceptibility 
of $(182(8)~{\rm MeV})^4$ and an instanton size $\sim 0.6$~fm. 
In the presence of light staggered fermions and across the chiral transition,
the susceptibility drops in a manner consistent with the quenched case,
and the instanton size changes little. No significant formation of 
bound instanton-antiinstanton pairs is observed, in particular not along the 
Euclidean time direction for $T > T_c$.
\end{abstract}

\maketitle

 
Here, we extend to SU(3) our earlier study of topological properties
of the SU(2) vacuum using improved cooling \cite{su2}.
The topology of lattice gauge fields is obscured by lattice artifacts
(dislocations). We remove them by cooling. To avoid losing at the same
time physical, large-scale topological structures, we cool with a
highly improved action. This action provides a small energy barrier
against the decay of instantons of size $\rho \gtap 2.3 a$, thus guaranteeing
stability of the measured topological charge under {\em arbitrary}
amounts of cooling. Furthermore, it gives a scale-invariant instanton
action, up to terms ${\cal O}(a/\rho)^6$, ensuring good stability of the
instanton size under cooling. Using 5 planar loops (1x1, 2x2, 3x3, 1x2, 1x3),
we construct a 1-parameter family of such classically improved actions,
then tune this parameter for maximum scale-invariance. We use for SU(3)
the same action as for SU(2), with the same tuning coefficient.
Because an SU(3) instanton is constructed by embedding an SU(2) instanton,
and because cooling preserves this embedding, there is no need to
retune the SU(3) action. 


\section{Pure gauge}

We have applied improved cooling to SU(3) gauge theory at $\beta=5.85~(12^4)$
and $6.0~(16^4)$, on 120 configurations each, separated by 50 MC sweeps
(in a 4:1 mixture of OR and PHB). The MC time history of the topological
charge $Q$ at $\beta=6$, shown in Fig.1, reveals autocorrelations already
known to plague full QCD simulations. 
Since topological sectors are disconnected from each other in continuum
field theory, energy barriers of height diverging as $a \rightarrow 0$ must
be present in the lattice theory (full or quenched QCD). The autocorrelation
of $Q$ is a clear indication that we are measuring a relevant physical
quantity: any contribution to $Q$ of lattice artifacts (eg. dislocations)
would appear as additional noise and quickly wash out the autocorrelation.

\begin{figure}[tbh]
\begin{center}
\mbox{
\epsfxsize=7cm
\epsffile[130 275 553 750]{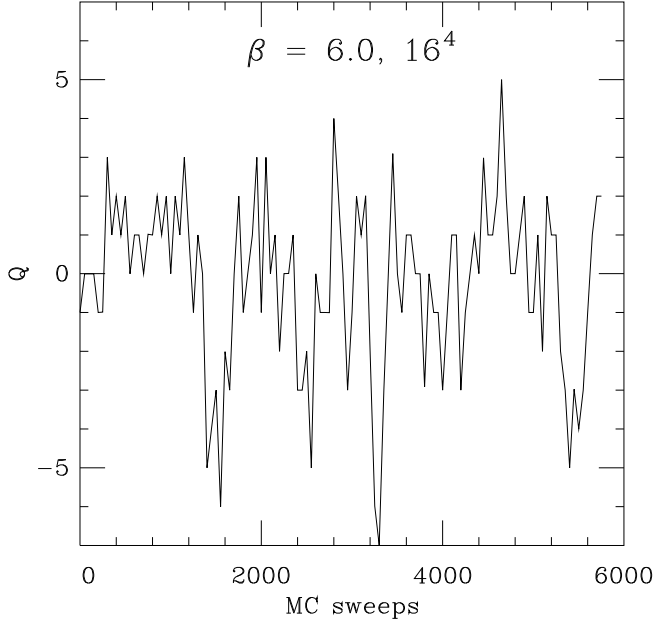}}
\end{center}
\vspace*{-1.0cm}
\caption{Evolution of the topological charge in quenched Monte Carlo.}
\end{figure}

The resulting topological susceptibility $\chi$ is $(185(7)~{\rm MeV})^4$
$[\beta=5.85, a=0.134~{\rm fm}]$ or $(182(8)~{\rm MeV})^4$ $[\beta=6, a=0.1~{\rm fm}]$,
in good agreement with Ref.\cite{Pisa}.
The size distribution we measure (see Fig.2) scales well with $\beta$.
A slight shift to larger sizes is observed under cooling, and fits of
the form $\rho^6 \exp(-(\rho/w)^c)$ are somewhat cooling-dependent 
(see figure). However the peak of the distribution stays around $0.6$ fm.
Just like for SU(2) where the distribution peaked around $0.43$ fm,
we observe a significantly larger size than expected from instanton
liquid models
(cf. also talk by D. Smith).
Together with the observed instanton density, it makes a dilute
description of the instanton fluid rather questionable.
\nopagebreak

\begin{figure}[tbh]
\begin{center}
\mbox{
\epsfxsize=7cm
\epsffile[88 174 572 698]{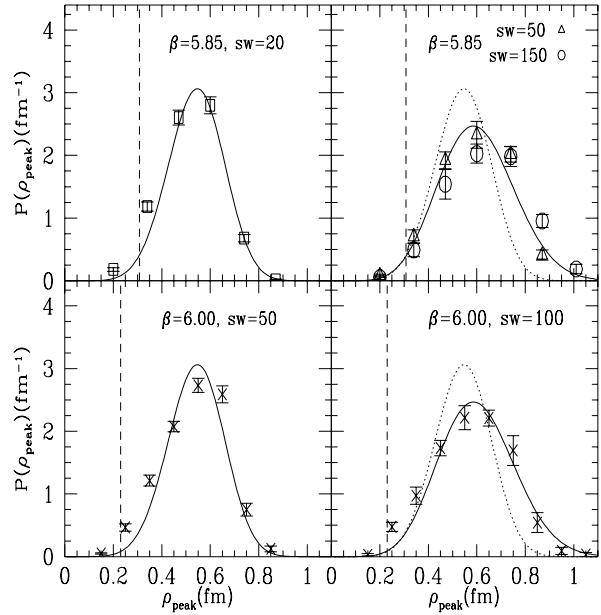}}
\end{center}
\vspace*{-0.7cm}
\caption{Size distribution for quenched SU(3).}
\end{figure}

\section{Full QCD}

We have analyzed configurations generously provided by the MILC 
collaboration \cite{MILC}:
$24^3 \times 12$, 2 flavors of staggered fermions, $ma = 0.008$. 
The dynamical quarks are thus rather light, with $m_\pi / m_\rho \sim 0.6$.
Our sample consists of 31, 43, 20 configurations respectively at 
$\beta =  5.65$, $5.725$ and $5.85$, corresponding to 
$\frac{T}{T_c} \approx 0.95, 1.05$ and $1.25$.
Ergodicity problems are glaring at $\beta = 5.65$, since we measure
$\langle Q\rangle = -1.68 \pm 0.33$ over the complete simulation run by MILC. At higher
$\beta$, $\langle Q^2\rangle$ is smaller, making the lack of
topological tunneling less important.
Ignoring these problems, one can infer the following values for the
topological susceptibility, based on the measured values of $\langle Q^2\rangle$:
$(134(10)~{\rm MeV})^4$, $(102(5)~{\rm MeV})^4$ and $0$ at the 3 temperatures
(at $\beta = 5.85$ all 20 configurations have $Q = 0$).
Thus the dynamical quarks suppress $\chi$ at the lowest temperature as
expected; further suppression  as a function of $T/T_c$ appears 
consistent with the pure gauge case \cite{Pisa}.

The extraction of instanton sizes should be modified from the $T = 0$ case:
in the continuum at finite $T$, an instanton is distorted by periodic
$T-$images into a ``caloron'', and these distortions become important
for our case where $\rho T \sim {\cal O}(1/2)$. Ignoring these effects
for the time being, the typical instanton size observed
remains $\sim 0.6$ fm, with little dependence on temperature.

\begin{figure*}[t]
\begin{center}
\mbox{
\epsfxsize=15cm
  \epsffile[30 30 1081 769]{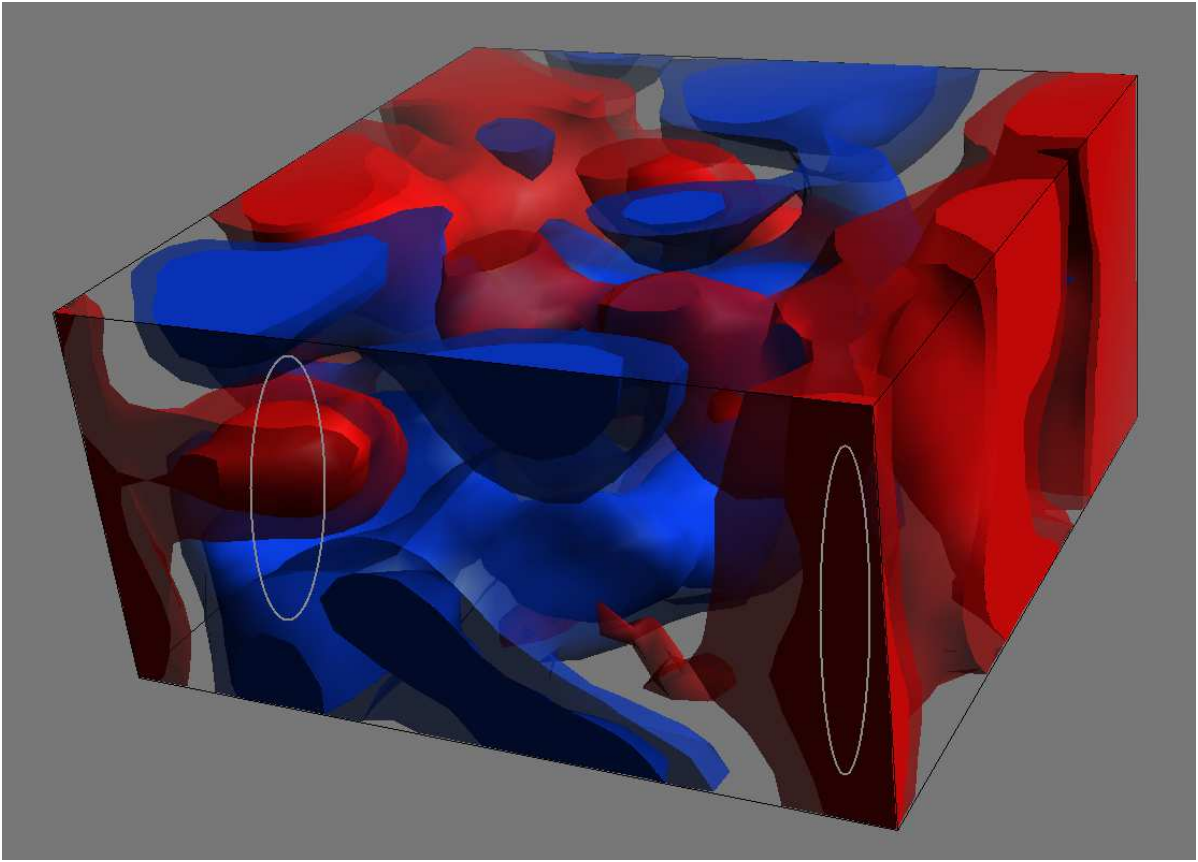}}
\end{center}
\vspace*{-0.7cm}
\caption{Isosurfaces of topological charge density ($\beta=5.85$, i.e.
$T/T_c \approx 1.25$) after 20 cooling sweeps. 
The left ellipse indicates the presence of an instanton-antiinstanton
pair oriented along Euclidean time (the short direction), as predicted by the instanton liquid
model. The right one shows a single static instanton. The solid and 
transparent surfaces correspond to a density of $\pm 84$ and $\pm34$
$(8 \pi^2)/{\rm fm}^4$ respectively.}
\vspace*{-0.5cm}
\end{figure*}

Instantons have been argued \cite{Shuryak} to play a determining role in 
the chiral phase transition. The fermionic determinant induces an attractive
force between objects of opposite topological charge, which thus tend
to form dipoles; according to the instanton-liquid model 
at $T \sim T_c$ the attraction increases and 
free charges become rare, which explains the suppression of topological 
fluctuations; moreover these dipoles would align predominantly along Euclidean time.
To check these predictions, we have measured the correlation
$<q(\vec{0},0) q(\vec{x},t)>$ of the topological charge density as a function
of the space and time separations $\vec{x}$ and $t$.
This observable can be measured at any level of cooling or with no cooling
at all, and it bypasses any possible bias in the identification of instantons,
which becomes delicate 
and ambiguous at short cooling.
Some negative correlation is observable at short distance under 
short cooling; 
the space-time anisotropy of these negative correlations is usually weak, 
${\cal O}(10\%)$ or less.
Reflection positivity enforces negative correlations 
starting at some distance in the absence of
cooling, so it is hard to decide whether what we measure is really caused
by genuine instanton-antiinstanton pairs.
To gain some qualitative insight, we monitored isosurfaces of topological
charge density $q(\vec{x},t) = constant$. Fig.3
shows that two effects are in place: 
some alternating $\pm$ charges at the same spatial location, in accordance
with the instanton liquid scenario (left of figure); and for the most part,
large single static charges, expected in other models (eg. \cite{Smit_vdS})
(right of figure). One may argue that the quarks in the present simulation
are still too massive ($m_\pi / m_\rho \sim 0.6$) for the predictions of
the instanton liquid scenario to apply.

\end{document}